\documentclass[twocolumn,showpacs,superscriptaddress,10pt]{revtex4-1} 
\usepackage{amsmath,amssymb,graphicx}

\newcommand{\rmi}{{\rm i}}

\begin{document}

\title{Composite device for interfacing an array of atoms with a single nanophotonic cavity mode}
\author{Mark Sadgrove}
\affiliation{Research Institute of Electrical Communications, Tohoku University, Sendai 980-8577, Japan}
\author{Kali P. Nayak}
\affiliation{Center for Photonic Innovations, The University of Electro-Communications, 1-5-1 Chofugaoka, Tokyo, Japan}

\begin{abstract}
We propose a method of trapping atoms in arrays near to the surface of a composite nanophotonic device with optimal coupling to a single cavity mode.
The device, comprised of a nanofiber mounted on a grating, allows the formation of periodic optical trapping potentials near to the nanofiber surface along with
 a high cooperativity nanofiber cavity. We model the device analytically and find good agreement with numerical simulations. 
We numerically demonstrate that for an experimentally realistic device, an array of traps can be formed
whose centers coincide with the antinodes of a single cavity mode, guaranteeing optimal coupling to the cavity. Additionally, we simulate a trap suitable for a single 
atom within 100 nm of the fiber surface, potentially allowing larger coupling to the nanofiber than found using typical guided mode trapping techniques. 
\end{abstract}

\maketitle

\section{Introduction}
Interactions of atoms with the electromagnetic field near micro and nano-scale dielectric structures 
is currently a topic of great interest~\cite{Kimble,Valhalla}. On the one hand, atoms in the vicinity of 
microscopic resonators can interact strongly with photons in the resonator leading to quantum
information applications such as single photon switching~\cite{Toroids}. On the other hand, recent advances 
have been made in trapping arrays of atoms near the surface of nanofibers using guided modes~\cite{LeKien1,ArnoTrap,KimbleNanof,MarkOIST} 
for which large optical densities are realizable -- a feature which is also conducive to 
quantum information as well as more general quantum optics applications. Trapping of single atoms near to nanostructure based cavities 
has also seen recent advances~\cite{LukinNWG, KimblePhC,Aoki}.
\begin{figure}[h!]
\centering
\includegraphics[width=\linewidth]{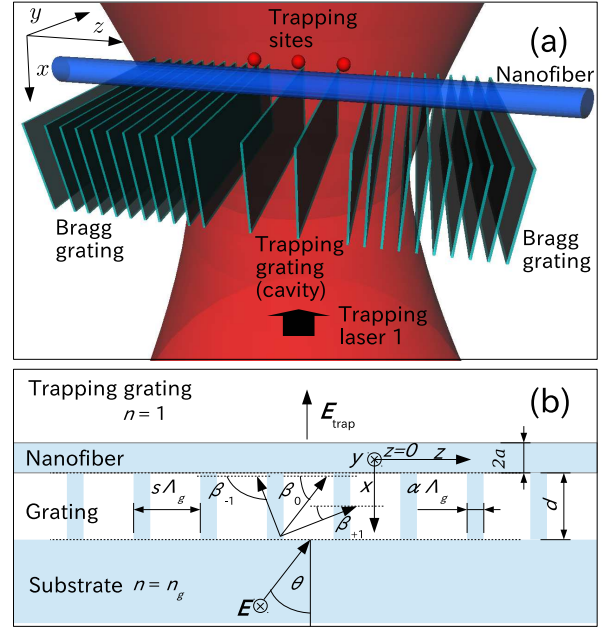}
\caption{\label{fig:setup} (a) Depiction of the trapping scheme showing a nanofiber mounted on a grating illuminated
from below (i.e. from $x=+\infty$ towards $x=-\infty$) by a focused laser beam. This conceptual diagram omits the grating substrate for clarity.
Scattering of the diffracted beam by the fiber produces an array of traps near the fiber surface.
(b) Schematic diagram showing the regions of importance for the theoretical derivation of the trapping field. A fiber of radius
$a$ is mounted on a grating of period $s\Lambda_g$, for integer $s$, depth $d$ and slat width $\alpha\Lambda_g$, where $\alpha$ is the grating
duty cycle. A $y$-polarized
plane wave is incident on the grating from the substrate side at an angle $\theta$. The grating diffracts the incident wave into
three orders with angles $\beta_{-1},\beta_0$ and $\beta_{+1}$ with respect to the fiber axis. These plane waves are then
scattered by the fiber giving rise to a trapping field $\mathbf{E}_{\rm trap}$. }
\end{figure}

Nonetheless, a number of challenges remain in this area. One problem is the trapping of atoms in optimum positions
for coupling to nanowaveguide-based cavities~\cite{Aoki}. In standard, freespace cavity QED experiments, this problem has already been 
solved with applications to atomic self organization~\cite{Barret} and atomic spin squeezing~\cite{KasevichOL,KasevichNat}, but these methods do not translate well to the nanowaveguide case. 
Another challenge is the vector light shift imparted on the atoms when they are trapped 
using the guided mode of a nanowaveguide~\cite{FamEJPhysD}. As an additional matter, the observation of collective excitation
effects such as superradiance using nanofibers~\cite{FamSuper} would be greatly simplified if the spacing of a trapping array 
could be decoupled from the wavelength of light used to create it.  

Here we describe a scheme to trap atoms in arrays near to the surface of a composite nanostructure which can potentially solve 
these problems as well as provide new features compared to the aforementioned studies.
The device we consider here is a nanostructure comprised of a 
nanofiber mounted on a grating. Similar devices have recently been experimentally investigated for the purposes of enhancing the coupling
of a quantum emitter to the guided modes of an optical nanofiber~\cite{OurGrating,OurPRL,OurOL}.
Here, we will show that by illuminating such a device, it is possible to create periodic arrays of
trapping potentials near to the nanofiber surface as depicted in Fig.~\ref{fig:setup}(a).
 
The basic principle of the device is that a longer period grating (referred to here as a \emph{trapping grating}) between two Bragg gratings on a nanostructured silica substrate can serve the 
dual purposes of 1) enhancing emission into the integrated nanofiber by acting as a cavity and 2) acting as a first order grating for trapping light 
which illuminates the device. The principle advantages of the proposed device are as follows:
a) An array of traps can be formed along the nanofiber surface, with the number of traps and their period 
determined by the design of the nano-structured substrate, rather than the wavelength of the trapping light.
A major application of this, as we will show later, is trapping devices where the trapping sites are \emph{geometrically guaranteed} to
have their centers aligned to antinodes of a cavity mode of the device.
b) Unlike trapping schemes which use the evanescent tails of the guided modes, in our scheme no trapping light is present in
the guided mode, which should lead to improved signal-to-noise ratios for detection of photons emitted into the guided mode.
c) Unlike trapping arrays formed by the guided modes of the nanofiber, the traps formed by this technique have linear rather than elliptical
polarization. This means the potentially decohering effect of vector light shifts~\cite{KimbleNFNJP,FamEJPhysD} is not present for our device.
d) The composite nature of the device has a number of benefits as given in~\cite{OurOL}. Of particular importance in this instance is the separation of the nanowaveguide
function and the index modulation function into separate devices. This allows for much greater diffraction efficiencies into the $\pm1$st order compared to what could be 
achieved with structures patterned directly onto the waveguide, leading to relatively larger trapping depths along the fiber axis.

The paper proceeds in the following way. First we consider the trapping potential formed when a plane wave is incident on 
the system with an infinitely long trapping grating and no Bragg-grating region, as depicted in Fig.~\ref{fig:setup}(b).
This system may be analysed both numerically and using an  analytical model which allows a crosscheck of the numerically calculated potential. 
Having established the validity of our numerics, we then go on to simulate the full device, including the Bragg mirror regions, using FDTD simulations.

\section{The device}
The device is depicted in a to-scale conceptual image in Fig.~\ref{fig:setup}(a) and in schematic form in 
 Fig.~\ref{fig:setup}(b). (Note that the grating substrate is omitted in Fig.~\ref{fig:setup}(a) for clarity). An optical nanofiber of radius $a$ is mounted 
on a nano-structured substrate and illuminated from the grating side (single-illumination configuration)
or from both sides (dual-illumination configuration) by a red-detuned laser beam with a vacuum wavelength $\lambda$. 
The corresponding wave number of the trapping laser is $k=2\pi/\lambda$ and the trapping laser frequency is $\omega=kc$, 
where $c$ is the speed of light in the vacuum.

The nano-structured substrate is divided into two regions. 
To either side of the center, a Bragg grating region is present. The two Bragg gratings, which have period $\Lambda_g$
 act as mirrors with respect to the fundamental mode of the nanofiber~\cite{OurGrating}. 
The central region is the trapping grating (shown in Fig.~\ref{fig:setup}(b))
with period $\lambda < s\Lambda_g < 2\lambda$ for some real number $s$, i.e., the trapping grating is a first order grating relative 
to the incident trapping light. We characterize the length of the trapping region by counting the number
of regions $N_t$ of width $s\Lambda_g$. In this paper, $N_t$ is always an odd number.
We note that with respect to the Bragg mirrors, the central trapping grating region may be considered to constitute a cavity of total width $N_t s\Lambda_g$. 
This cavity region acts to enhance coupling of spontaneous emission to the fiber guided modes.

In all regions of the nano-structured substrate, the grating slats have depth $d=2\;\mu$m and width $w = \alpha\Lambda_g=50$ nm, where $\alpha$ is the Bragg grating duty cycle.
As shown in Fig.~\ref{fig:setup}(b), light incident on the device at angle $\theta$ is diffracted into the $0$th and $\pm1$st orders at angles $\beta_0$ and $\beta_{\pm1}$ respectively.
In the remainder of the paper, we will take $\theta=0$ implying that the light is normally incident on the grating substrate.

We define our axes so that $z=0$ is always in the middle of the trapping grating region and $x=0$ corresponds to the fiber center as shown in Fig.~\ref{fig:setup}(b).
The unit vectors $\mathbf{e}_x$, $\mathbf{e}_y$ and $\mathbf{e}_z$ are defined to lie along the $x$, $y$ and $z$ axes respectively. 

In the analytical treatment which follows below, we will only consider the case of normally incident, $y$-polarized
trapping light which is red-detuned from the atomic resonance. The effect of changing the incident angle, polarization 
or detuning will be discussed in the discussion section.

\section{Evaluation of the trapping potential}

\subsection{Analytical treatment}

\subsubsection{Field created by illumination of the composite device}

By applying the theory of a planewave scattered from a dielectric grating~\cite{Knop} together with scattering theory for a sub-wavelength dimension cylinder~\cite{BohrenHuffman} 
the field outside the device for the incident $y$-polarized field can be found. A detailed derivation is given in the Appendix. 
To summarize the theoretical approach: We make the approximation that evanescent orders of the grating may be ignored, thereby truncating the series which gives the output field.
We also assume that the field around the nanofiber can be found by taking the field at the output of the grating as the input field to the scattering problem involving the nanofiber .

To find the field around the fiber we need to find the scattered field due to the $0$th and $\pm1$st order plane waves. To apply the standard scattering formalism,
we first evaluate the angles $\beta_\ell$ where $\ell\in\{-1,0,1\}$ as shown in Fig.~\ref{fig:setup}(b). In the 
present paper, we will consider only the case where the trapping beam is at normal incidence to the
grating, giving $\beta_0=\pi/2$. The other angles are given by $\beta_{\pm 1} = \delta_{-1,\pm 1}\pi + \tan^{-1}(t_{\pm 1}/p_{\pm 1})$, where $\delta$ is the Kronecker delta function,
$t_{\pm 1}$ is the wavenumber of the $x$ component of the diffracted field and $p_{\pm 1}$ is the wavenumber of the $z$ component of the diffracted field.

The field for $x<-a$ is then given by
\begin{eqnarray}
\label{eq:EtrapI}
\mathbf{E}_{{\rm trap},I} &=& E_{\rm g} + \sum_{\ell=-1}^1 T_\ell\exp(\rmi k\sin(\beta_\ell) a) \sum_{n=-\infty}^\infty \frac{(-{\rm i})^n}{k\sin(\beta_\ell)}\times\nonumber\\
& & [\rmi a_n(\beta_\ell)\mathbf{M}_n(x,y,z) + b_n(\beta_\ell)\mathbf{N}_n(x,y,z)],
\end{eqnarray}
where $E_g$ is the field due to the grating in the absence of the nanofiber (See Appendix), $T_\ell$ is the transmission coefficient of the $\ell$th diffraction 
order, $\mathbf{M}_n$ and $\mathbf{N}_n$ are $n$th order cylindrical harmonics whose coefficients $a_n$ and $b_n$ respectively depend on the incident angles of the transmitted orders of the grating. 
These coefficients are defined fully in the Appendix, where a brief review of scattering theory for a dielectric cylinder is given for the case considered here.

\subsubsection{Dual illumination}
In the dual illumination scheme, a second laser illuminates the device from the fiber side.
To derive the trapping field in this case, we will assume that the only fields to be considered are those of the 
second trapping laser itself, and the scattered field caused by the second trapping laser being incident on the nanofiber.
That is, we ignore any reflections from the grating or, indeed, any effects due to the grating at all. Because of the low reflectivity of the grating at the wavelength considered here, this approximation is not a bad one. In particular, for the  
space $x<-a$, the assumption turns out to produce good agreement with FDTD simulation results. The incident field is 
given by
\begin{equation}
\label{eq:EiII}
\mathbf{E}_{{\rm i},II} = \exp(\rmi\Omega)\exp(\rmi kx)\mathbf{e}_y,
\end{equation}
where $\Omega$ is the relative phase between trapping laser 1 and trapping laser 2 which we will
assume is adjustable. The scattered field is then
\begin{equation}
\label{eq:EsII}
\mathbf{E}_{{\rm s},II} = \exp(\rmi\Omega)\sum_{n=-\infty}^{\infty} \frac{(-{\rm i})^n}{k\sin(\beta_\ell)} \rmi a_n\mathbf{M}_n(-x,y,z),
\end{equation} 
where the sign of $x$ in the argument of the spherical harmonics $\mathbf{M}_n$ has been flipped to account for the 
fact that the incident wave is coming from the negative $x$ direction, and the terms in $\mathbf{N}_n$ are zero
due to the normal incidence of the second trapping beam. Finally, we may write the field due to illumination
by both trapping beams 1 and 2 as
\begin{equation}
\label{eq:DualEtrap}
\mathbf{E}_{{\rm trap},II} = \mathbf{E}_{{\rm trap},I} + \mathbf{E}_{{\rm i},II} + \mathbf{E}_{{\rm s},II}.
\end{equation}

Note that in what follows, we will use $\mathbf{E}_{\rm trap}$ to denote either the trapping potential for single or dual illumination. The meaning of
$\mathbf{E}_{\rm trap}$ will always be clear from the context in which it is used.

\subsubsection{Calculation of the trapping potential}
To calculate the trapping potential experienced by $^{133}$Cs atoms, we follow closely the formalism of Ref.~\cite{LeKien1}.
The optical trapping potential may be calculated as
\begin{equation}
\label{eq:optpot}
U_{{\rm opt}} = -\frac{1}{4}\alpha|E_{\rm{trap}}|^2,
\end{equation}
where
\begin{equation}
\alpha(\omega) = 2\pi\varepsilon_0c^3\sum_j\frac{g_j}{g_a}\frac{A_{ja}(1-\omega^2/\omega_{ja}^2)}{(\omega_{ja}^2-\omega^2)^2+\gamma^2_{ja}\omega^2}.
\end{equation}
The index $j$ runs over different transitions of cesium that contribute to the trapping potential. For our
purposes, we will include the four dominant lines of the atom~\cite{LeKien1} corresponding to wavelengths
$\lambda_{1a} = 852.113$nm, $\lambda_{2a} = 894.347$nm, $\lambda_{3a} = 455.528$nm and  $\lambda_{4a} = 459.317$nm, where $\lambda_{ja}=2\pi c/\omega_{ja}$.
The transition strengths for these lines are $A_{1a}=3.276\times10^7$s$^{-1}$, $A_{2a}=2.87\times10^7$s$^{-1}$,
$A_{3a}=1.88\times10^6$s$^{-1}$ and $A_{4a}=8\times10^5$s$^{-1}$. The statistical weights of each transition are
$g_1=4$, $g_2=2$, $g_3=4$, $g_4=2$ and the ground state has weight $g_a=2$. Finally, we define the 
linewidths $\gamma_{ja}=\sum_{j'}A_{jj'}$, where $j'$ denotes a lower level $|j'\rangle$. 
However for the detuning of the trapping light considered here, the contribution of the terms $(\gamma_{ja}\omega)^2$ to $\alpha$ are negligible.

For convenience, all of our analytical and simulation results assume an electric field amplitude of 1. We scale the potential
to provide trap depth predictions for a given optical power as follows. For simplicity, we use the waist region of a circular
Gaussian beam as a reference. We assume a power $P_0$ and a beam 
waist radius of $r_0$. Then the peak intensity in the beam is given by $I_{\rm peak} = 2P_0/(\pi r_0^2)$. 
The scaling factor for the optical part of the trapping potential is then $\chi=I_{\rm peak}/(\varepsilon_0 c)$, where $\varepsilon_0$ is the vacuum permittivity.

Next, we consider the contribution of the van der Waals potential at the fiber surface to the trapping potential.
Here we will use the simple flat-surface, bulk-medium van der Waals potential experienced by a Cs atom close to a silica surface which is given in mK by
\begin{equation}
\label{eq:VdW}
U_{\rm vdW} = - \frac{4.1\times10^{-5}}{(r-a)^3},
\end{equation} 
where $r=\sqrt{x^2+y^2}$ is measured in $\mu$m.
This flat potential approaches the Van der Waals potential due to a silica fiber when the distance from the fiber surface approaches zero~\cite{LeKien1}.
The van der Waals potential due to the grating is neglected as the trapping potential minima always lie between two grating slats, i.e., $\sim500$ nm away from the nearest
slat for the parameters used in this paper. This is sufficiently far that the contribution from the grating slats to the van der Waals potential is negligible.

The total trapping potential is the sum of the optical and Van der Waals potentials:
\begin{equation}
\label{eq:trappot}
U = \chi U_{\rm opt} + U_{\rm vdW},
\end{equation}
where it is understood that the factor $\chi$ is only necessary to scale the optical potential if the analytical results and/or simulations have used incident fields of unity amplitude.

\subsection{Comparison of theory and finite-difference time-domain numerical calculations}
\label{sec:theorynumcomp}
\begin{figure}[h!]
\centering
\includegraphics[width=\linewidth]{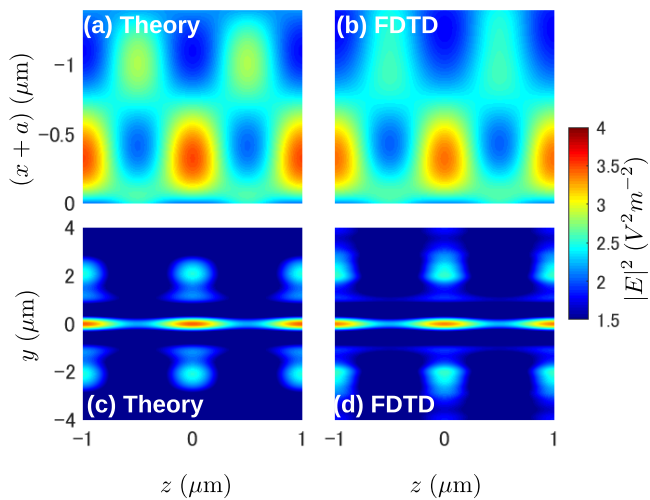}
\caption{\label{fig:intens} Comparison of intensities $I$ as predicted by theory and
FDTD simulations. (a) and (b) show $|\mathbf{E}_{\rm trap}|^2$ in the $x-z$ plane for $y = 0$ as given by theory and FDTD numerics respectively.
In (c) and (d), $|\mathbf{E}_{\rm trap}|^2$ from theory and simulations respectively is shown in the $y-z$ plane for $x+a=0.35\;\mu$m. The fiber radius was $a=0.3\;\mu$m and $\Lambda_g=0.350\;\mu$m 
with $s=3$.}
\end{figure}
\begin{figure}[h!]
\centering
\includegraphics[width=\linewidth]{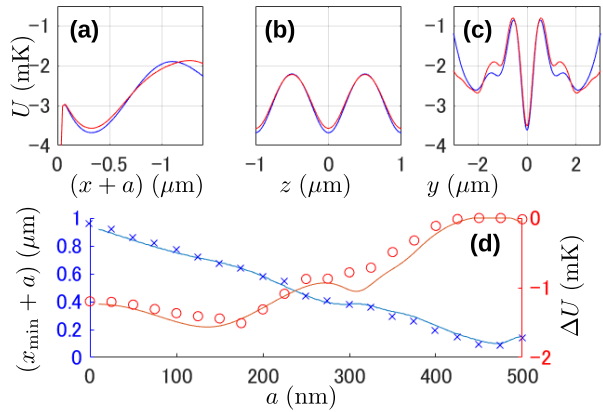}
\caption{\label{fig:pot} Trapping potential produced near the nanofiber surface for (a) $y=0,z=0$, (b) $x+a=0.35\;\mu$m , $z=0$ and (c) $x=0.35\;\mu$m, $y=0$
respectively, corresponding to Figs.~\ref{fig:intens}(a-c). In all cases, solid blue lines show the theoretical prediction while red lines show the predictions from
FDTD simulations. (d) shows how the calculated position of the trapping minimum (blue line) and trap depth (red line) vary as a function
of the fiber radius. The blue crosses show the trap position and the red circles show the trap depth
given by the FDTD simulation in each case. The optical power was set to 250 mW.}
\end{figure}
We simulated the electromagnetic field in the system shown in Fig.~\ref{fig:setup}(b) for an incident plane wave using the finite difference time domain (FDTD) method (Lumerical Inc.).
Here, and throughout the remainder of this paper, we assume an optical power of 250 mW per beam, a beam spot size of 10 $\mu$m and a trapping light wavelength of 937 nm (the red detuned magic wavelength for $^{133}$Cs)
in order to calculate the trapping potential generated by the intensity distribution.

Figure~\ref{fig:intens} shows a comparison between squared electric field amplitudes as calculated from Eq.~\ref{eq:EtrapI} and by using FDTD numerical simulations.
We see good agreement between the squared field profiles in the $x-z$ plane (Figs.~\ref{fig:intens}(a) and (b)) and the $y-z$ plane (Figs.~\ref{fig:intens}(c) and (d)) 
both in the qualitative shape of the field pattern and the intensity. In particular, Figs.~\ref{fig:intens}(a) and (b) show the generation of a periodic array of intensity maxima close to the
nanofiber surface. Figs.~\ref{fig:intens} (c) and (d) show that along the axis perpendicular to the fiber, for $x$ set to
the position of the intensity maximum in the $x-z$ plane, a strong intensity maximum coincides with the center of the fiber.
Away from the center is an area of low intensity, and for larger $y$, the diffraction pattern due to the grating is restored 
since the field scattered by the fiber tends to zero in this region.

Figures~\ref{fig:pot}(a-c) show the corresponding trapping potential as a function of $x$, $z$ and $y$ respectively, 
as calculated from Eq.~\ref{eq:trappot} and from the FDTD data in all cases assuming a power of $P=250$ mW. We see that potential depths of mK order are achievable 
in all directions for these conditions, although the effect of the Van der Waals potential significantly lowers the potential
depth along the $x$-axis. Fig.~\ref{fig:pot}(d) shows how the trapping depth $\Delta U$ and the trap minimum position $x_{\rm min}$ on the $x$-axis vary as the fiber 
radius $a$ is varied. It may be seen that increasing the nanofiber radius leads to the trapping potential moving closer to the nanofiber surface, 
with the trapping potential depth $|\Delta U|$ decreasing due to the effect of the Van der Waals potential. 
It may be seen that for trapping minima positions less that $\sim200$ nm the potential depth becomes vanishing.

\begin{table}[]
\centering

\begin{tabular}{l|ll|}
\cline{2-3}
                       &    Trap freq.                   & (kHz) \\ \hline
\multicolumn{1}{|l|}{Axis} & \multicolumn{1}{l|}{Theory} & FDTD  \\ \hline
\multicolumn{1}{|l|}{$x$} & \multicolumn{1}{l|}{816} & 770       \\ \hline
\multicolumn{1}{|l|}{$y$} & \multicolumn{1}{l|}{1270} & 1320      \\ \hline
\multicolumn{1}{|l|}{$z$} & \multicolumn{1}{l|}{982} & 939      \\ \hline
\end{tabular}
\caption{\label{tab:SingleIllum} Theoretical and FDTD calculated trapping frequencies for single illumination.}
\end{table}
In Table~\ref{tab:SingleIllum}, we show the trapping frequencies along each axis for the theoretically and FDTD calculated trapping potentials 
shown in Figs.~\ref{fig:pot}(a-c). In all cases we find fair agreement between theoretically and FDTD calculated values.

\begin{figure}[h!]
\centering
\includegraphics[width=\linewidth]{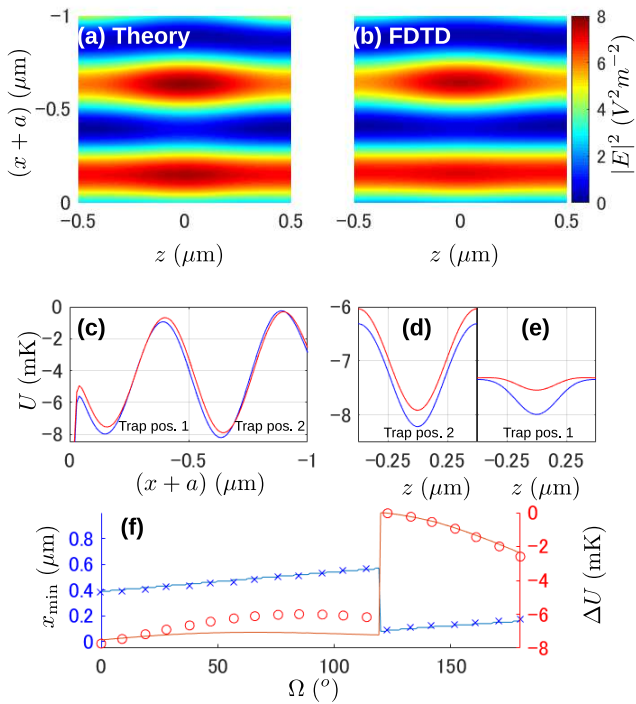}
\caption{\label{fig:dualillum}  (a) and (b) show $|\mathbf{E}_{\rm trap}|^2$ in the $x-z$ plane for $y = 0$ as given by Eq.~\ref{eq:DualEtrap} and 
FDTD numerics respectively for a relative phase $\Omega=180^o$.
(c) shows the trapping potential produced near the fiber surface as a function of
$x$ for $y=0, z=0$ in the case of dual illumination.
 Solid blue lines show the theoretical prediction while red lines show the predictions from
FDTD simulations. The fiber radius is $a=0.3\;\mu$m, the trapping grating period is $1\;\mu$m and the relative phase $\Omega$ is fixed at 180$^o$. 
(d) and (e) show the trapping potential as a function of $z$ at $y=0$ for the $x$ positions indicated.
(f) shows the position $x_{\rm min}$ of the trapping potential minimum closest to the fiber surface (blue line) and trap depth (red line) as a function
of the relative phase $\Omega$ between the two trapping beams. The blue crosses show the trapping potential minimum position given by FDTD simulations,
while red circles show FDTD simulation trap depths. In all cases, the optical power was $P = 250$ mW in both beams.}
\end{figure}

We will now show that a dual illumination scheme can produce trapping potentials much closer to the fiber-surface while maintaining potential depths of 
order $|\Delta U|\approx500\;\mu$K. Because the introduction of the counter-propagating beam principally affects the potential shape along the $x$-axis, we focus
on the $x$-dependence of the potential in what follows.

Figure~\ref{fig:dualillum} (a) shows the squared electric field amplitude as given by Eq.~\ref{eq:DualEtrap}, while 
the results of FDTD simulations are shown in Fig.~\ref{fig:dualillum}(b). Unlike the single illumination case, because a standing wave pattern is formed, a number of trapping 
potentials are formed in the $x-z$ plane. We will focus on the traps which are nearest and next-nearest to the fiber surface as shown in 
Figs.~\ref{fig:dualillum}(a,b). Figure~\ref{fig:dualillum}(c) shows the $x$-dependence of the trapping potential  for $z=0,y=0$ and a relative phase of $\Omega=180^o$ between trapping beams 1 and 2.
Of note is that the position of the trapping potential minimum closest to the nanofiber surface (Trap pos. 1) is only $x_{\rm min}\approx100$ nm. The $x$-dependent trapping potential at this position has
a depth of  $\sim 2$ mK. Additionally, a much deeper trap is found about 600 nm from the fiber surface (Trap pos. 2).

Figures~\ref{fig:dualillum} (d) and (e) show the $z$ dependence of the trapping potential at the $x$-axis trapping centers furtherest and closest to the fiber surface 
respectively. As seen in Fig.~\ref{fig:dualillum} (e), there is a large difference in the trap depth predicted by the  theoretical model and the FDTD data when the trap is close to the 
fiber surface. This might be due to the presence of evanescent fields from higher order grating modes which are explicitly neglected in our  theoretical model.
Fig.~\ref{fig:dualillum} (e) thus illustrates the limits of the simple model in quantitatively predicting the potential depth along the $z$ axis. 
Figure~\ref{fig:dualillum}(f) shows the dependence of the trapping potential depth $\Delta U$ and the position of the trapping minimum 
$x_{\rm min}$ as the phase between the two trapping lasers is varied. Up to about $\Omega=120^o$, the smaller
trapping potential is not separated from the fiber surface and the closest trapping potential is formed by the second intensity
maximum at $\sim400$ nm from the fiber surface. For phases greater than $\Omega=120^o$, the 
trapping potential closer to the fiber surface becomes distinct and the values of $x_{\rm min}$ and $\Delta U$ jump to the new levels associated with this trapping potential as seen in Fig.~\ref{fig:dualillum}(f).
The ability to smoothly control the position and depth of the trapping potential by varying the phase between the two lasers may be useful for loading of the the trapping sites.

\begin{table}[]
\centering
\begin{tabular}{l|llll|}
\cline{2-5}
                       &                       & Trap freq.                &     (kHz)              &             \\ \cline{2-5} 
                       &              Trap         & \multicolumn{1}{l|}{  pos. 1  } &      Trap                 &  pos. 2 \\ \hline
\multicolumn{1}{|l|}{Axis} & \multicolumn{1}{l|}{Theory} & \multicolumn{1}{l|}{FDTD}     & \multicolumn{1}{l|}{Theory} &   FDTD          \\ \hline
\multicolumn{1}{|l|}{$x$} & \multicolumn{1}{l|}{4100} & \multicolumn{1}{l|}{4000}     & \multicolumn{1}{l|}{4320} &      4270       \\ \hline
\multicolumn{1}{|l|}{$z$} & \multicolumn{1}{l|}{809} & \multicolumn{1}{l|}{539}     & \multicolumn{1}{l|}{1140} & 1100             \\ \hline
\end{tabular}
\caption{\label{tab:DualIllum} Theoretical and FDTD calculated trapping frequencies for dual illumination.}

\end{table}
In Table~\ref{tab:DualIllum}, we show the trapping frequencies along the $x$ axis for the theoretically and FDTD calculated trapping potentials 
shown in Figs.~\ref{fig:dualillum}(c-e) for both trapping position 1 (closest to the nanofiber) and trapping position 2.
Results for the $z$ axis are also shown for comparison, but as expected, the values are little changed from the single-illumination case, when the trap is sufficiently far from the fiber surface.
We find fair agreement between theoretical and FDTD calculated values in all cases, except in the case of the $z$-dependence of the potential closest to the fiber, for reasons noted above.
 
\section{Numerical results for experimentally realistic structures}
\label{sec:real}
\begin{figure}[htb]
\centering
\includegraphics[width=\linewidth]{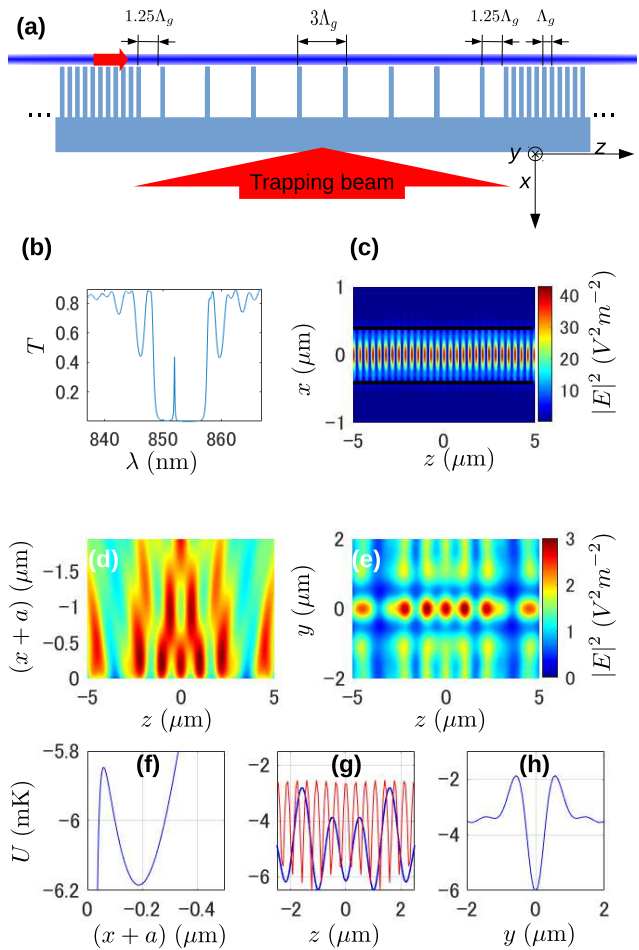}
\caption{\label{fig:gaussillumsingle} (a) Schematic diagram of the device under single illumination. 
The horizontal red arrow indicates the guided mode of the nanofiber.  (b) Transmission spectrum of the fundamental mode of the nanofiber. (c)
Intensity distribution of the cavity mode shown in (b).
(d) $x-z$ and (e) $y-z$ plane squared field amplitudes. In (f)-(h), the trapping potentials associated with 
(d) and (e) are shown. In (g), the cavity mode intensity is overlayed in arbitrary units (red curve). The fiber radius was 300 nm.}
\end{figure}
\begin{figure}[h!]
\centering
\includegraphics[width=\linewidth]{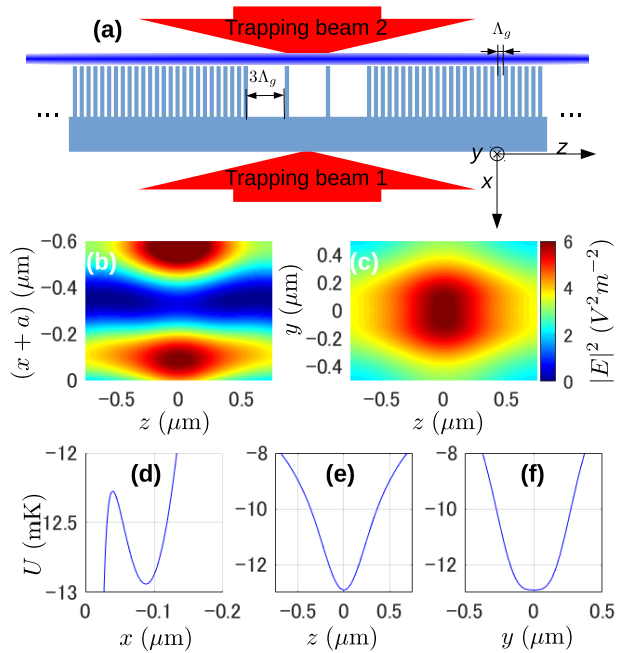}
\caption{\label{fig:gaussillumdouble} (a) Schematic diagram of the device under dual illumination. Intensities in the (b) $x-z$ and (c) $y-z$ planes. In (d)-(f), the trapping potentials associated with 
these intensity profiles are shown. The fiber radius was 300 nm.}
\end{figure}
In the following section, we simulate the trapping potential created by experimentally realistic structures for which the trapping grating is of finite length
and a Bragg mirror structure exists to either side of the trapping grating.
All parameters are the same as in Section~\ref{sec:theorynumcomp}.

In order that the positions of the traps coincide with antinodes of the cavity mode, a modification to the structure of the device is necessary.
We wish to make the cavity mode antinodes appear exactly in the middle of the trapping grating slats so that they coincide 
with the center of the optical trap. To achieve this, it is necessary that while the trapping grating period is an integer multiple of $\Lambda_g$ as before,
the overall length of the cavity region should be 
given by $[(2n+1)+1/2]\Lambda_g$ for integer $n\geq0$ (which is an odd multiple of $\Lambda_g/2$). We achieve this condition by adding two shorter grating regions of length
$1.25\Lambda_g$ at each end of the trapping grating region, as illustrated in Fig.~\ref{fig:gaussillumsingle}(a). 

We set $N_t=7$ for the trapping grating, nanofiber diameter $2a=600$ nm, grating period $\Lambda_g=350$ nm, 
and $s=3$, giving a trapping grating period of $3\Lambda_g=1050$ nm. The transmission spectrum through the nanofiber for the fundamental mode is shown in Fig.~\ref{fig:gaussillumsingle}(b).
Note that a single cavity mode arises near the middle of the photonic stop band. 
The associated intensity distribution at the cavity resonance frequency is shown in Fig.~\ref{fig:gaussillumsingle}(c).
Typical Q-values for composite cavities such as this device have been measured to be $\sim 2000$
with associated cooperativities (equivalent to the Purcell factor in the regime considered here) of $\sim 10$~\cite{OurPRL}.

Figures~\ref{fig:gaussillumsingle}(d) and (e) show the squared field pattern in the $z-x$ and $z-y$ planes respectively and 
Figs.~\ref{fig:gaussillumsingle}(f-h) show the associated trapping potentials formed  along the $x$, $z$ and $y$ axes respectively.
As seen in Fig.~\ref{fig:gaussillumsingle}(f), the trapping potential is formed about 200 nm from the fiber surface and the trap depth along the $x$-axis is $\sim 300\;\mu$K. 
Although 7 trapping cavities are present, Fig.~\ref{fig:gaussillumsingle}(g) shows that only 5 distinct trapping potentials, with trap depths of $\sim 2$ mK, are seen along the z axis due to
the decreasing intensity of the Gaussian beam away from $z=0$ along with the weaker diffraction pattern at the edge of the
trapping grating.

Figure~\ref{fig:gaussillumsingle}(g) also shows the intensity distribution of the cavity mode
overlayed as a red line. This figure demonstrates an important property of our deivce: namely that the trapping potential minima can be made to coincide almost perfectly
with cavity antinodes by careful design of the grating structure. Indeed, the slight difference between the cavity anti-node positions and the trapping minima positions is mainly due
to edge effects of the relatively short $N_t=7$ trapping grating. We would expect even better coincidence between the trapping sites and cavity antinodes as $N_t$ is increased.
Figure~\ref{fig:gaussillumsingle}(h) shows that the trapping potential measured in the $y$ direction has the largest trap depth of $\approx 4$ mK.

The second case we consider is that of trapping a single atom to achieve large coupling efficiency with the nanofiber.
We note that $N_t=3$ is the minimum number of trapping cavities necessary to achieve a trapping minimum in the middle of the trapping cavity.
Therefore we choose $N_t=3$, and choose a double illumination scheme which can achieve trapping nearer to the nanofiber surface compared with single illumination
as discussed in Section~\ref{sec:theorynumcomp}.

Figure~\ref{fig:gaussillumdouble}  shows the trapping potential formed by dual illumination in the central trapping region where $-0.5<z<0.5\;\mu$m for an optical power of 250 mW in each beam.
The relative phase between the trapping beams was $\Omega=133^o$.
Figures.~\ref{fig:gaussillumdouble}(b) and (c) show the squared field pattern in the $z-x$ and $z-y$ planes respectively.
As seen in Fig.~\ref{fig:gaussillumdouble}(d),  a trapping minimum is formed very close ($\sim 80$ nm) to the fiber surface with a trapping depth of about 1 mK.
The trapping potential along the $x$-axis is much tighter than in the case of single illumination.
Figs.~\ref{fig:gaussillumdouble}(e) and (f) show that the trap depth along the $z$ and $y$ axes is $\approx 5$ mK.
 \begin{table}[]
\centering

\begin{tabular}{l|ll|}
\cline{2-3}
                       &    Trap freq.                   & (kHz) \\ \hline
\multicolumn{1}{|l|}{Axis} & \multicolumn{1}{l|}{Single Illum.} & Dual Illum.  \\ \hline
\multicolumn{1}{|l|}{$x$} & \multicolumn{1}{l|}{1180} & 5300      \\ \hline
\multicolumn{1}{|l|}{$y$} & \multicolumn{1}{l|}{1440} & 1460      \\ \hline
\multicolumn{1}{|l|}{$z$} & \multicolumn{1}{l|}{1164} & 1480      \\ \hline
\end{tabular}
\caption{\label{tab:SingleDual}Trapping frequencies for FDTD calculated single and dual illumination trapping configurations.}
\end{table}
Table~\ref{tab:SingleDual} compares the trapping frequencies for FDTD calculated trapping potentials in the cases of single and dual illumination as shown in Figs.~\ref{fig:gaussillumsingle}(e-g) and~\ref{fig:gaussillumdouble}(d-f)
respectively. As expected, we find essentially no change along the $y$-axis, and the change along the $z$-axis is small. However the $x$ axis value is more than quadrupled due to the tighter trap created 
by the standing wave as well as the closer proximity of the trap to the fiber surface.

\section{Discussion and conclusion}

We now consider aspects of our system not discussed in the preceding sections. 
First the question arises as to our choice of parameters for the trapping light. Regarding blue detuned trapping schemes, because the focusing effect of the nanofiber
determines the distance of the trapping potential minimum from the fiber surface and the focus is always an intensity maximum, trapping sites tend to be further from the nanofiber
in the blue detuned case. We also calculated the potential at the blue-detuned magic wavelength of 686 nm, 
but we found that red-detuned trapping light produced traps closer to the nanofiber surface.

As for the polarization of the trapping beam, in this work we used exclusively $y$ polarized light. This polarization is parallel to the grating slats, so that the 
output polarization of the grating is also $y-$polarized. On the other hand, $y-$ polarized illumination of the nanofiber produces a field which includes localized $x-$polarized components at
azimuthal angles of $\sim 45^o$ on the fiber surface. Nonetheless, we found that the polarization of the trapping field is also $y$-polarized near the trapping minima to a good approximation,
with less than $1\%$ $x-$polarization at the trap center, and slightly less than $10\%$ $x-$ polarized at 300 nm away from the trap center in the $y-$direction (corresponding to a temperature of $\sim1$ mK).
Additionally, we found that illuminating the device with $z$-polarized light did not produce trapping potentials due to the mixture of polarizations resulting after diffraction 
by the grating.

Next, we note that in the case of non-normal incidence, a diffraction pattern (and thus a trap) is  only produced if the incident angle $|\theta_{\rm in}|<5^o$.
This is because the gratings considered here are strictly first order gratings. This means that the device must be precisely aligned in order to produce a trapping potential.

Additionally, here we have principally considered the trapping sites closest to the fiber surface. 
This is justified by the exponential scaling of the coupling between a dipole emitter and the fiber guided mode, atoms trapped in more distant 
trapping sites have negligible coupling to the nanofiber~\cite{LukinNWG}. 

Another issue to consider in the present scheme is the possibility that large light intensities might affect the integrity of the very thin silica slats on the grating.
However, the slats are at intensity nodes of the trapping field for the cases considered here. Additionally, the trapping intensity produced by the interference of diffracted orders of 
the grating is only of order twice the input intensity. For these reasons the thin grating slats are not in general exposed to large field intensities.

Lastly, we consider issues related to the tuning of the cavity. Illumination of the device may cause thermal expansion and thus drifting of the cavity mode
from its design wavelength. We note, however, that much of the optical power for circular beam illumination is unused, due to the thinness of the nanofiber. 
If, for example, we used an elliptical beam with a $z$-axis radius of 10 $\mu$m as before but an $y$-axis radius of only 1 $\mu$m, the optical power required to 
achieve the same intensity at the device surface would be only $10\%$ of that necessary in the case of a circular beam (i.e. $\sim 25$ mW). 
In this way heating of the silica substrate would be drastically reduced. If tuning of the cavity resonance is required, it can be achieved by slightly rotating the nanofiber
relative to the grating slats (increasing the resonance wavelength) or by lifting the fiber slightly off the grating substrate (reducing the resonance wavelength).

In conclusion, we have analytically and numerically investigated trapping atoms close to the surface of a composite nanooptical device consisting  
of an optical nanofiber mounted on a grating. Illumination of the device from the grating side produces an array of traps with separations from the nanofiber surface down to 
200 nm. Adding illumination from the fiber side  can produce trapping potentials within 100 nm of the nanofiber surface. Bragg gratings to either side of the trapping grating create
a cavity enhancing the coupling of spontaneous emission into the nanofiber guided modes, and by appropriately designing the grating structure, arrays of atoms may be trapped at positions
which correspond to antinodes of the cavity mode. Because optical nanofibers offer automatic coupling to standard fiber optical networks, and due to the advantages of the present scheme
over existing nanofiber trapping schemes, we expect that it may be useful in a variety of situations. These include the realization of collectively enhanced scattering
into the nanofiber guided modes from an array of atoms, through to quantum information applications where distant atoms must be coupled to the same cavity mode.

MS acknowledges support from a grant-in-aid for scientific research (Grant no. 15K13544) from the Japan Society for the Promotion of Science (JSPS).
KPN acknowledges support from a grant-in-aid for scientific research (Grant no. 15H05462) from the Japan Society for the Promotion of Science (JSPS).
The authors thank K. Hakuta for his scientific input at the early stages of this research and for his support. We also thank F. Le Kien for discussions regarding the van der Waals 
potential, M. Morinaga for advice regarding the scattering problem, and R. Yalla for discussions related to the numerical simulations.

\appendix

\section{Numerical simulation protocols}
For comparison with theoretical results, we performed FDTD simulations using plane wave sources in our simulations. 
For scaling the potential, we choose a beam waist radius of $\omega_0=10\;\mu$m, an optical power of $P_0 = 250$ mW, and a wavelength of $\lambda = 937$ nm.
In the dual illumination case, we assumed beam recycling was possible so that the power in both forward and backward propagating beams was set to $P_0$.
In FDTD simulations of 
periodic systems using plane  waves, care must be taken in interpreting the results since artifacts of the FDTD method can form regular periodic intensity maxima of a 
similar nature to the true trapping maxima we wish to find. We used periodic boundary conditions for these simulations to ensure that no edge-effects due to the 
FDTD boundary conditions were present. We also checked that removing all structures in the simulation gave rise to 
a flat intensity profile, and that removing the grating structure led to the disappearance of all $z$-periodic structures in the output intensity pattern. 
Once these checks had been passed, we could be confident that any local intensity maxima in the simulations corresponded to the physical nature of the scattered field 
rather than numerical artifacts. Comparison with the  analytical results given in the previous sections also
functioned as a further cross-check of our numerical results.

We note that in the case of dual illumination, the relative phase $\Omega$ is just the phase of the second
beam as set in our FDTD simulations. It is neccesary to match the phase between the theoretical trapping potential and the FDTD calculated trapping potential for a single data set
before the FDTD and theory results can be compared for arbitrary phases.

In the simulations for the case of realistic experimental structures, we illuminated the device with a Gaussian beam. We used absorbing rather than periodic boundary conditions, 
and it was necessary to make the FDTD simulation region large enough that the Gaussian beam tails had decayed effectively to zero by the edges of the simulation region to avoid edge effects
such as reflection from the absorbing boundary.  However, computer memory limits the mesh density which can be achieved in these large simulation regions. 
We used linear interpolation to reconstruct the results on a finer grid to improve the detail visible in the Figures. As before, we checked the validity of our simulations by removing all structures 
from the simulation region and checking that only a Gaussian intensity profile remained in such cases.

\section{Field due to a plane wave incident on a dielectric grating}
\label{app1}
In our analytical derivation of the trapping potential we will use the axes and variables defined in Fig.~\ref{fig:setup}(b). 
In particular, we assume a grating structure with period $s\Lambda_g$, where $\Lambda_g$ is the Bragg grating period, $s$ is an integer chosen so that the trapping
grating is a first order grating. The grating structure has a depth $d$ and the grating slats have thickness $\alpha\Lambda_g$, where $\alpha$ is the duty cycle.
The grating is assumed to have refractive index $n=1.45$ at the trapping light wavelength, and for simplicity the grating structure is assumed to 
be fabricated on top of a silica substrate which has much larger dimensions than any of the other scales in the problem. (In typical experiments the substrate has been of 
size 15 mm x 5 mm x 2 mm~\cite{OurGrating}). The axes are defined as shown in Fig.~\ref{fig:setup}(b) and the unit vectors $\mathbf{e}_x$, $\mathbf{e}_y$, and $\mathbf{e}_z$
are aligned with the $x$, $y$, and $z$ axes respectively.

A plane wave is assumed to be incident on the structure originating from the $x=+\infty$ direction at an angle $\theta$ as shown in Fig.~1(b) of our manuscript. The grating
diffracts the incident wave into $0$th and $\pm 1$st orders with angles $\beta_0$ and $\beta_{\pm 1}$ respectively.
In what follows, standard harmonic time dependence $\exp(-i\omega t)$ is assumed for all fields and is not explicitly shown.

We first consider the electric field at the output of the grating \emph{in the absence of the nanofiber}. This problem
was solved by Knop in~\cite{Knop} but due to a typographical error, the result given was ambiguous. For clarity, we therefore reproduce the results of~\cite{Knop} below
giving the correct fromula which we have checked by rederiving the results.
For an incident field
\begin{equation}
\mathbf{E}_{{\rm i},I} = \exp[-{\rm i}kx]\mathbf{e}_y
\end{equation}
 the field outside the grating is given by
\begin{equation}
\label{eq:Enofiber}
\mathbf{E}_{\rm g}(x,z) = \sum_\ell T_\ell \exp(\rmi t_\ell a)\exp[\rmi(p_\ell z - t_\ell x)]\mathbf{e}_y,
\end{equation}
where
\begin{equation}
\label{eq:pdef}
p_\ell = 2\pi\ell/\Lambda_g + k\sin\theta,
\end{equation}
is the $z-$axis wavenumber and
\begin{equation}
\label{eq:tdef}
t_\ell = \left\{ \begin{array}{lr} \sqrt{k^2-p_\ell^2} & k\geq|p_\ell| \\
\rmi\sqrt{p_\ell^2-k^2}& k\leq|p_\ell| \end{array} \right.,
\end{equation}
is the $x-$axis wavenumber for diffraction order $\ell$. The coefficients $T_\ell$ give the (complex) amplitude of each diffracted component.

Although they do not contribute to the trapping potential, we also define the $x-$axis wavenumbers of the plane wave components
reflected at the grating interface which are given by
\begin{equation}
\label{eq:rdef}
r_\ell = \left\{ \begin{array}{lr} \sqrt{n_0^2k^2-p_\ell^2} & n_0k\geq|p_\ell| \\
\rmi\sqrt{p_\ell^2-n_0^2k^2}& n_0k\leq|p_\ell|. \end{array} \right.
\end{equation}
Also, note that the phase term $\exp(\rmi t_\ell a)$, which does not appear in~\cite{Knop}, is necessary to retard the 
phase of the solution since the grating output is offset from the origin along the $x$-axis by $a$ (the output of the grating 
is at $x=0$ in~\cite{Knop}).

The coefficients $T_\ell$ of the different plane wave components are given by 
\begin{equation}
\label{eq:Tsol}
\mathbf{T} = \hat{U}^{-1}\mathbf{Y}, 
\end{equation} 
where $\hat{U}$ is a matrix with components
\begin{eqnarray}
\label{eq:Ucomp}
 {\hat{U}}_{l,m} & = & (r_l + t_m)\sum_n (\hat{E})_{l,n}(\hat{E}^{-1})_{n,m}\cos(g_n d) \nonumber \\
& & -\rmi\sum_ n \left(g_n + \frac{r_lt_m}{g_n}\right)(\hat{E})_{l,n}(\hat{E}^{-1})_{n,m}\sin(g_n d),\nonumber\\
& & 
\end{eqnarray}
where $\hat{E}$ is a matrix of eigenvectors and $g_n$ are the eigenvalues of the eigenvalue equation 
\begin{equation}
\label{eq:eve1}
\hat{A} \mathbf{E} = g^2 \mathbf{E},
\end{equation}
where $\hat{A}_{\eta,\ell} = k_0^2\alpha_{\eta-\ell} - \delta_{\ell,\eta}p_\ell^2$, and $\alpha_\ell$ is the $\ell$th
Fourier coefficient of the refractive index profile of the grating~\cite{Knop}.
The vector $\mathbf{Y}$ is made up of zeros except for a single element at $\ell=0$ which has the value
\begin{equation}
\label{eq:yell0}
(\mathbf{Y})_{\ell=0} = 2r_0\exp(-\rmi r_0 d).
\end{equation} 

It is well known that the numerical inversion of $\hat{U}$
required to calculate $T$ is unstable when significant numbers of evanescent diffraction orders are included in the 
calculation~\cite{KnopImprovedNumericsPaper}. In this paper, we will only consider first order gratings, i.e. all but
the $\pm 1$th and $0$th diffraction orders are evanescent. As noted in~\cite{Knop}, a simple way to avoid the problems
due to inversion of near-singular matrices is simply to truncate the eigenvalue problem. Therefore we  approximate $T$ as a length three vector 
and matrices $\hat{U}$, $\hat{E}$ and $\hat{A}$ are taken to have size $3\times 3$.
This means that we neglect all evanescent orders of the grating which is a good approximation when the distance from the grating is more than the wavelength
of the trapping light.
This approximation is justified by the good agreement it produces when compared to the results of FDTD numerical simulations.

\section{Review of the solution for a planewave incident at oblique incidence on a dielectric cylinder}
\label{app2}
We now review the analytical form of the field created when a plane wave is incident on a dielectric cylinder at oblique incidence with angle $\beta$.
The treatment follows exactly that of Ref.~\cite{BohrenHuffman} and is included here only for convenience, not as a new result of our study. The following results are valid for the case where 
the propagation vector $\mathbf{k}$ of the incident wave lies in the $x-z$ plane and the polarization is parallel to 
the $y$ axis.

We first define cylindrical coordinates for our system. The azimuthal coordinate $\phi$ is given 
by $\phi = \tan^{-1}(y/x)$, and the radial coordinate is given by $r = \sqrt{x^2+y^2}$. For the cylindrical
unit vectors, we define $\mathbf{e}_r=\cos\phi\mathbf{e}_x + \sin\phi\mathbf{e}_y$, the radial unit vector,
and $\mathbf{e}_\phi=-\cos\phi\mathbf{e}_x + \sin\phi\mathbf{e}_y$, the azimuthal unit vector.

To proceed, we introduce the vector cylindrical harmonics which are given by the following 
formulae:
\begin{widetext}
\begin{equation}
\mathbf{M}_n(x,y,z) = k\sqrt{1-\cos^2\beta}\left(\rmi n\frac{H_n}{\rho}\mathbf{e}_r - H'_n(\rho)\mathbf{e}_\phi\right)\exp[\rmi(n\phi - k\cos(\beta)z)],
\end{equation}
\begin{equation}
\mathbf{N}_n(x,y,z) = -\sqrt{1-\cos^2\beta}k\cos\beta\left(\rmi H'_n(\rho)\mathbf{e}_r - n\frac{H_n(\rho)}{\rho}\mathbf{e}_\phi + k\sqrt{1-\cos^2\beta} H_n(\rho)\mathbf{e}_z \right)\exp[\rmi(n\phi - k\cos(\beta)z)],
\end{equation}
\end{widetext}
where $\rho = r\sqrt{k^2-h^2}$, $h=-k\cos(\beta)$ and $H_n$ is the $n$th order Hankel function. Note that although the arguments to the 
cylindrical harmonics are given as cartesian coordinates, calculations take place in polar coordinates.

The field scattered by the cylinder can now be written as
\begin{equation}
\label{eq:Escat}
\mathbf{E}_s = \sum_{n=-\infty}^{\infty} \frac{(-{\rm i})^n}{k\sin(\beta)} [ \rmi a_n\mathbf{M}_n + b_n\mathbf{N}_n],
\end{equation} 
where the coefficients $a_n$ and $b_n$ are defined as follows.
\begin{equation}
a_n = - \frac{A_nV_n - \rmi C_nD_n}{W_nV_n + \rmi D_n^2},
\end{equation}
\begin{equation}
b_n = -\rmi \frac{C_nW_n + A_nD_n}{W_nV_n + \rmi D_n^2}.
\end{equation}

The $n$ dependent coefficients have the following defintions:
\begin{equation}
A_n = \rmi\xi[\xi J'_n(\eta)J_n(\eta) - \eta J_n(\eta)J'_n(\eta)],
\end{equation}
\begin{equation}
C_n = n\cos(\beta)\eta J'_n(\eta)J_n(\xi)(\xi^2/\eta^2 -1),
\end{equation}
\begin{equation}
D_n = n\cos(\beta)\eta J_n(\eta)H_n(\eta)(\xi^2/\eta^2 -1),
\end{equation}
\begin{equation}
V_n = \xi [n_f^2\xi J'(\eta)H_n(\xi) - \eta J_n(\eta)H_n'(\xi)],
\end{equation}
\begin{equation}
W_n = \rmi\xi [\eta\xi J'(\eta)H_n'(\xi) - \eta J_n'(\eta)H_n(\xi)],
\end{equation}
where $\xi = x\sin\beta$, and $\eta = ka\sqrt{n_f^2-\cos^2\beta}$.

Finally, the total field is found by adding the incident field $\mathbf{E}_{\rm i}$ to the scattered field, i.e.
\begin{equation}
\label{eq:Etot}
\mathbf{E}_{\rm tot.} = \mathbf{E}_{\rm i} + \mathbf{E}_{\rm s}. 
\end{equation}  

\section{Formula for the electric field outside composite device where $x<-a$}

We can now write down the formula for the electric field outside the device as used in our manuscript. We will approximate this field by
the output field of the grating as scattered by the nanofiber. Noting that $a_n$ and $b_n$ depend on the incident angle $\beta$, we write $a_n(\beta)$ and $b_n(\beta)$.
Then, by applying Eq.~\ref{eq:Etot} along with Eq.~\ref{eq:Escat} for each diffraction order of the grating, we can write the field outside the device in the region $x<-a$
for the case of single illumination as 
\begin{eqnarray}
\mathbf{E}_{{\rm trap}} &=& E_{\rm g} + \sum_{\ell=-1}^1 T_\ell\exp(\rmi k\sin(\beta_\ell) a) \sum_{n=-\infty}^\infty \frac{(-{\rm i})^n}{k\sin(\beta_\ell)}\times\nonumber\\
& & [\rmi a_n(\beta_\ell)\mathbf{M}_n(x,y,z) + b_n(\beta_\ell)\mathbf{N}_n(x,y,z)].
\end{eqnarray}


\begin{thebibliography}{99}
\bibitem{Kimble} H.J. Kimble, ``The quantum internet," Nature \textbf{453}, 1023-1030 (2008).
\bibitem{Valhalla} K.J. Vahala, ``Optical microcavities," Nature \textbf{424}, 839-346 (2003). 
\bibitem{Toroids}  T. Aoki, B. Dayan, E. Wilcut, W. P. Bowen, A. S. Parkins, T. J. Kippenberg, K. J. Vahala, and H. J. Kimble, 
``Observation of strong coupling between one atom and a monolithic microresonator," Nature 442, 671-674 (2006).
\bibitem{LeKien1} F. Le Kien, V. I. Balykin, and K. Hakuta, ``Atom trap and waveguide using a two-color evanescent light field around a subwavelength-diameter optical fiber," 
Phys. Rev. A \textbf{70}, 063403 (2004).
\bibitem{ArnoTrap} E. Vetsch, D. Reitz, G. Sagué, R. Schmidt, S. T. Dawkins, and A. Rauschenbeutel, 
``Optical Interface Created by Laser-Cooled Atoms Trapped in the Evanescent Field Surrounding an Optical Nanofiber," 
Phys. Rev. Lett. \textbf{104}, 203603 (2010).
\bibitem{KimbleNanof} A. Goban, K. S. Choi, D. J. Alton, D. Ding, C. Lacroûte, M. Pototschnig, T. Thiele, N. P. Stern, and H. J. Kimble, 
``Demonstration of a State-Insensitive, Compensated Nanofiber Trap," Phys. Rev. Lett. \textbf{109}, 033603 (2012).
\bibitem{MarkOIST} M. Daly, V.G. Truong, C.F. Phelan, K. Deasy and S. Nic Chormaic, ``Nanostructured optical nanofibres for atom trapping," New J. Phys. \textbf{16}, 053052 (2014).
\bibitem{LukinNWG} J. D. Thompson, T. G. Tiecke, N. P. de Leon, J. Feist, A. V. Akimov, M. Gullans, A. S. Zibrov, V. Vuletic, and M. D. Lukin, 
``Coupling a single trapped atom to a nanoscale optical cavity," Science \textbf{340} 1202-1205 (2013).
\bibitem{KimblePhC} C.-L, Hung, S.M. Meenehan, D.E. Chang, O. Painter, and H.J. Kimble, ``Trapped atoms in one-dimensional photonic crystals," New J. Phys \textbf{15}, 083026 (2013).
\bibitem{Aoki} S. Kato,,  and T. Aoki, ``Strong Coupling between a Trapped Single Atom and an All-Fiber Cavity," Phys. Rev. Lett. \textbf{115}, 093603 (2015).
\bibitem{Barret} K. J. Arnold, M. P. Baden, and M. D. Barrett, ``Self-Organization Threshold Scaling for Thermal Atoms Coupled to a Cavity," Phys. Rev. Lett. \textbf{109}, 153002 (2012).
\bibitem{KasevichOL}J. Lee, G. Vrijsen, I. Teper, O. Hosten, and M.A. Kasevich, ``Many-atom–cavity QED system with homogeneous atom–cavity coupling," Opt. Lett. \textbf{39}, 4005-4008 (2014).
\bibitem{KasevichNat} O. Hosten, N.J. Engelsen,	R. Krishnakumar, and Mark A. Kasevich, ``Measurement noise 100 times lower than the quantum-projection limit using entangled atoms," Nature \textbf{529}, 505-508 (2015).
\bibitem{FamEJPhysD}  F. Le Kien, P. Schneewies, and A. Rauschenbeutel, ``Dynamical polarizability of atoms in arbitrary light fields: general theory and application to cesium," Eur. Phys. J. D \textbf{67}, 92 (2013).
\bibitem{FamSuper} F. Le Kien, S. Dutta Gupta, K. P. Nayak, and K. Hakuta, ``Nanofiber-mediated radiative transfer between two distant atoms," Phys. Rev. A \textbf{72}, 063815 (2005).
\bibitem{OurGrating} M. Sadgrove, R. Yalla, K. P. Nayak, and K. Hakuta, ``Photonic crystal nanofiber using an external grating," Opt. Lett. \textbf{38}, 2542-2545 (2013).
\bibitem{OurPRL} R. Yalla, M. Sadgrove, K. P. Nayak, and K. Hakuta, ``Cavity Quantum Electrodynamics on a Nanofiber Using a Composite Photonic Crystal Cavity," Phys. Rev. Lett. \textbf{113}, 143601 (2014).
\bibitem{OurOL} J. Keloth, M. Sadgrove, R. Yalla, and K. Hakuta, ``Diameter measurement of optical nanofibers using a composite photonic crystal cavity," Opt. Lett., \textbf{40}, 4122-4125 (2015).
\bibitem{KimbleNFNJP} C. Lacroute, K.S. Choi, A. Goban, D.J. Alton, D. Ding, N.P. Stern, and H.J. Kimble, ``A state-insensitive, compensated nanofiber trap," New J. Phys. \textbf{14}, 023056 (2012).
\bibitem{Knop} K. Knop, J. Opt. Soc. Am. \textbf{68}, 1206 (1978).
\bibitem{BohrenHuffman} C. F. Bohren and D. R. Huffman, ``Absorption and Scattering of Light by Small Particles", Wiley-VCH (2004).
\bibitem{KnopImprovedNumericsPaper} M. G. Moharam, Eric B. Grann, and Drew A. Pommet, J. Opt. Soc. Am. \textbf{12}, 1068 (1995).
\end{thebibliography}
\end{document}